# Super FSR tunable optical microbubble resonator

M. Sumetsky[*)], Y. Dulashko, and R. S. Windeler
*OFS Laboratories, 19 Schoolhouse Road, Somerset, NJ 08873*
[*)] *Corresponding author: sumetski@ofsoptics.com*

An optical resonator is often called fully tunable if its tunable range exceeds the spectral interval that contains the resonances at all the characteristic modes of this resonator. For the high Q-factor spheroidal and toroidal microresonators, this interval coincides with the azimuthal free spectral range. In this Letter, we demonstrate the first mechanically fully tunable spheroidal microresonator created of a silica microbubble having a 100 micron order radius and a micron order wall thickness. The tunable bandwidth of this resonator is more than two times greater than its azimuthal free spectral range.

**OCIS codes:** (060.2340) Fiber optics components; (230.3990) Micro-optical devices; (140.4780) Optical resonators; (140.3948) Microcavity devices; (130.6010) Sensors.

A wide class of optical microresonators having the shape of spheres and spheroids [1-5], toroids [6,7], disks [8,9], bottles [10-12], etc. support the high Q-factor whispering gallery modes (WGMs) which are localized near the equator of these resonators as illustrated in Fig. 1(a). For the axially symmetric resonators, the wavenumber spectrum of these modes can be approximately defined by the equation

$$k_{mpq} \approx R_0^{-1}\left[ m + \Delta + 2^{-1/3} t_p m^{1/3} + \left(q + \tfrac{1}{2}\right)(R_0/R_1)^{1/2} \right], \quad (1)$$

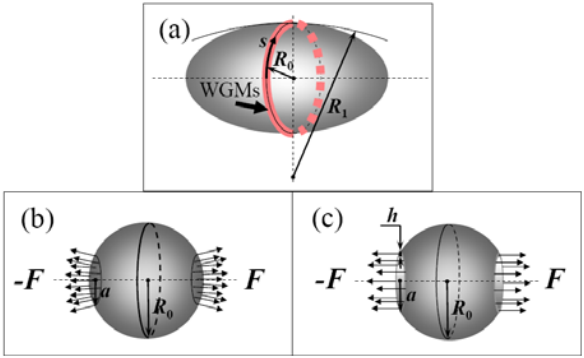

Fig. 1. (a) – Illustration of an optical microresonator. (b) – Bulk spherical microresonator with applied stretching force. (c) – Hollow spherical microresonator with applied stretching force.

where $k_{mpq} = 2\pi n_{eff}/\lambda_{mpq}$, $\lambda$ is the radiation wavelength, $n_{eff}$ is the effective refractive index of WGMs, $t_p$ are the roots of the Airy function, $R_0$ and $R_1$ are the radii of the microresonator surface along the equator and the axis of symmetry, respectively, and $\Delta \sim 1$ is a polarization-dependent shift having a relatively weak dependence on the WGM quantum numbers $m$, $p$, and $q$ (see, e.g., [2,10]). The azimuthal quantum number $m$ in Eq. (1) is large, $m \gg 1$. It corresponds to the exponential dependence $\exp[\pm ism/(2\pi R_0)]$ of a WGM from the coordinate $s$ along the equator. Integers $q$ and $p$ are responsible for the WGM behavior along the directions transversal to $s$. The numbers $p$ and $q$ are relatively small so that a WGM is localized near the equator. The quantum number $p$ counts the roots of a WGM along the radial direction, while the number $q$ is the number of roots along the axial direction. The WGM radial and axial dependencies can be approximately expressed through the Airy function and Hermite polynomial, respectively (see, e.g. [13]).

The spectrum defined by Eq. (1) is approximately periodic in $m$ so that the experimentally measured positions of resonance peaks are also approximately periodic with the period equal to the azimuthal free spectral range (FSR) $\Delta k_{FSR}^{(azim)} = R_0^{-1}$ corresponding to the wavelength FSR

$$\Delta\lambda_{FSR}^{(azim)} = (2\pi n_{eff} R_0)^{-1} \lambda^2. \quad (2)$$

More importantly, due to the exponential dependence $\exp[\pm ism/(2\pi R_0)]$ on $m$, which is uniform in amplitude and qualitatively similar for all WGM, it often happens that the measured spectral power distribution has the same period $\Delta\lambda_{FSR}^{(azim)}$. Thus, any spectral interval of length $\Delta\lambda_{FSR}^{(azim)}$ contains the full set of resonances corresponding to any type of WGMs with numbers $p$ and $q$ [14]. For this reason, a microresonator is called fully tunable if its tuning wavelength interval exceeds the FSR $\Delta\lambda_{FSR}^{(azim)}$ [3,12]. A resonance corresponding to any transversal WGM distribution of a fully tunable microresonator can be shifted to any predetermined laser wavelength.

The problem of creation of wide-range tunable microresonators received a substantial amount of attention [15-18]. In particular, creation of a fully tunable high Q-factor microsphere-type microresonator is of special interest [1-3,5,7-9,12]. For several applications (e.g., for tuning a microresonator to the frequencies of atomic and molecular transitions in cavity quantum electrodynamics [19-21]), the full tunability is important to adjust the modes possessing the strongest field confinement corresponding to the fundamental WGMs having $p = q = 0$. For other applications (e.g., for tunable narrow linewidth microlasers, optical filters, switches, and evanescent sensors [1-5]), a fully tunable microresonator provides remarkable functionality when any type of the existing WGM intensity distribution is available at any radiation wavelength.

Both thermal and mechanical tuning of high Q-factor WGM microresonators have been considered. For practical temperature range, thermal tuning of a microsphere [5] allows less tunability than mechanical tuning. While direct mechanical deformation of a microtoroid is problematic, Ref. [7] demonstrated a microtoroid resonator having the thermal tunability $\sim 0.35 \Delta\lambda_{FSR}^{(azim)}$. A mechanically tunable microsphere resonator was demonstrated both by squeezing [2] and stretching

[3,12]. The record mechanical tunability, $\sim 0.5 \Delta\lambda_{FSR}^{(azim)}$, was achieved in [3] for a microsphere having a 210 μm diameter. In [8,9], a LiNbO$_3$ disk WGM resonator with a macroscopic diameter 4.8 mm was fully tuned electrically. However, application of the approach [8,9] to achieve the full tunability for a much smaller microresonator, like considered in this Letter, requests an impractically large applied voltage. The authors of Ref. [12] experimentally considered a bottle microresonator [10] defined by the elongation condition $R_0 \ll R_1$. As follows from Eq. (1), for such a resonator, the spacing between the $q$-series of resonances, corresponding to the axial FSR $\Delta\lambda_{FSR}^{(axial)} = (2\pi n_{eff})^{-1}(R_0 R_1)^{-1/2}\lambda^2$, is smaller than that between the $m$-series, $\Delta\lambda_{FSR}^{(azim)} = (2\pi n_{eff} R_0)^{-1}\lambda^2$, by the factor $(R_1/R_0)^{1/2}$ and, therefore, is easier to accomplish. In [12], tuning between the resonances having $q=0$ and $q=1$ was demonstrated by stretching. To achieve the full tunability, bridging of the azimuthal FSR, $\Delta\lambda_{FSR}^{(azim)}$, with several $q$-series axial FSRs, $\Delta\lambda_{FSR}^{(axial)}$, was suggested. However, unlike the situation with the FSR $\Delta\lambda_{FSR}^{(azim)}$ for the $m$-series, the transversal structure of the WGMs corresponding to different $q$ is qualitatively different. Therefore, the problem of full tunability, i.e., ability to shift a microresonator to a resonance at an WGM with any predetermined transversal structure (e.g., to a fundamental transverse WGM having $p=q=0$) remained unsolved.

In this Letter, the problem of full tunability of a few hundred μm diameter high Q-factor microresonator is solved for the first time using the recently demonstrated silica microbubble resonator (MBR) [22]. An MBR considered in this Letter is a silica bubble with dimensions of the order of hundred microns and an extremely small wall thickness of the order of a micron. The MBR is created along a silica microcapillary by simultaneous local melting with a CO$_2$ laser and pressurizing. Generally, the method developed in [22] allows to fabricate small silica MBRs with diameters ranging from several hundreds to several μm. Similar to a silica microsphere resonator [1-5], an MBR is a high Q-factor resonator, which possesses series of WGMs localized near the equator. Eq. (1) can be applied to an MBR after modification of the $p$-dependent term by taking into account the microbubble wall thickness $h$. In particular, for $h$ smaller than the radiation wavelength $\lambda$, the $p$-series shrinks to $p=0$. An MBR exhibits several advantageous properties. For example, a small wall thickness of an MBR, which has the order of the radiation wavelength, allows one to use this resonator as a very responsive evanescent sensor of the microfluid or gas propagating inside the MBR. Besides, the thin wall makes an MBR much more responsive to the mechanical deformations, which can be used for strain and pressure measurements. For the same reason, a hollow MBR is more favorable for mechanical tuning than a similar bulk microresonator. The microcapillary pigtails, which are continuously integrated with an MBR, can be conveniently used for mechanical stretching or compressing of this resonator.

Prior to describing the results of our experiment, let us qualitatively compare the behavior of bulk and hollow WGM resonators under stretch. It is known that the major contribution into the spectral shift of microsphere-shaped microresonators is given by the variation $\Delta R$ of their equatorial radius $R_0$ [2,3]. Ignoring the relatively small effects of change in refractive index and in axial radius $R_1$, we determine the resonance wavelength shift as

$$\Delta\lambda \approx \Delta R \lambda / R_0 . \qquad (3)$$

From here and Eq. (2), the tunability figure of merit (FOM) is defined as

$$\text{FOM} = \Delta\lambda / \Delta\lambda_{FSR}^{(azim)} = 2\pi n_{eff} \Delta R / \lambda . \qquad (4)$$

To find $\Delta R$ for a given applied stretching force $F$, a spherical microresonator with simplified stretch conditions is considered (Fig. 1(b)). The bulk sphere with radius $R_0$ is assumed to be pulled by the radial pressure uniformly distributed along the surfaces with radius $a$ resulting in stretching force $F$. Using the analytical solution to this problem [23], we approximate the change $\Delta R_0$ of the equatorial radius $R_0$ in the form $\Delta R = (0.25 - 0.083\nu - 0.085 a^2/R_0^2)F/(ER_0)$, where $E$ and $\nu$ are the Young modulus and the Poisson ratio of the sphere material, respectively. For $a < 0.7 R_0$, this expression is valid with better than 1%. From here, a useful estimate for the radius variation can be obtained:

$$\Delta R \sim 0.2 F / (ER_0) \qquad (5)$$

Now, let us consider a model of a hollow sphere with wall thickness $h$, which is stretched with vertical force $F$ uniformly distributed along the aperture of radius $a$ (Fig. 1(c)). From the elastic shell theory [24] we find:

$$\Delta R = (1+\nu)F / (2\pi Eh) \sim 0.2 F / (Eh) , \qquad (6)$$

where in the last expression we used $\nu = 0.17$ for silica. Comparison of Eqs. (3), (5) and (6) yields the ratio of forces $F_{bulk}$ and $F_{hollow}$ necessary to achieve the same resonant wavelength shift in a bulk and in a hollow sphere:

$$F_{bulk} / F_{hollow} = R_0 / h , \qquad (7)$$

For typical MBR parameters, this ratio is $\sim 100$ [22]. The maximum $\Delta\lambda$ is limited by the ultimate strength of silica $p_{max} \sim 1-5$ GPa [25]. For a bulk sphere, the corresponding maximum force and wavelength shift are $F_{bulk}^{(max)} = \pi p_{max} a^2$ and

$$\Delta\lambda_{bulk}^{(max)} \sim 0.2\pi p_{max}(\lambda / E)(a / R_0)^2 , \qquad (8)$$

respectively. For a hollow sphere, we have $F_{hollow}^{(max)} = 2\pi p_{max} R_0 h$ and

$$\Delta\lambda_{hollow}^{(max)} \sim 0.4\pi p_{max}(\lambda / E) . \qquad (9)$$

Thus, the ratio of the largest possible resonance shifts, which can be achieved for a hollow and a bulk sphere, is

$$\Delta\lambda_{hollow}^{(max)} / \Delta\lambda_{bulk}^{(max)} \sim 2R_0^2 / a^2 . \qquad (10)$$

For example, for the parameters of the microresonator considered in [3], $R_0 = 105$ μm, $a = 45$ μm, Eq. (7) predicts ten times larger tunability of the hollow sphere compared to that of a bulk one. From Eqs. (8) and (9), for $\lambda = 0.8$ μm considered in [3] and $p_{max} \sim 1$ GPa, the maximum tunability of the bulk microresonator investigated in [3] is $\Delta\lambda_{max}^{bulk} \sim 1$ nm, while a similar hollow resonator has $\Delta\lambda_{max}^{hollow} \sim 10$ nm. More accurate comparison of hollow and bulk microresonator models can be performed by numerical simulations.

The MBR used in our experiment is shown in Fig. 2(a). It is fabricated by the CO$_2$ laser heating and pressurizing method described in [22]. The MBR is created along a microcapillary with radius $a = 60$ μm and has the spheroidal shape with equatorial radius $R_0 = 110$ μm and axial radius $R_1 = 160$ μm. The measured microcapillary wall thickness is 5.5 μm (Fig. 2(b)). The MBR wall thickness is $R_0/a$ times smaller, i.e., $h \approx 3$ μm. The azimuthal FSR of this resonator at $\lambda \approx 1.55$ μm considered in our experiment is $\Delta\lambda_{FSR}^{(azim)} \approx 2.5$ nm. For this wavelength and $p_{max} \sim 1$ GPa, Eq. (9) estimates the tunable range of this resonator as $\Delta\lambda_{max}^{hollow} \sim 20$ nm, which is 6.6 times larger than that for a similar bulk resonator. In addition, the stretching force, which is needed to achieve the same wavelength shift, is ~ 36 times smaller.

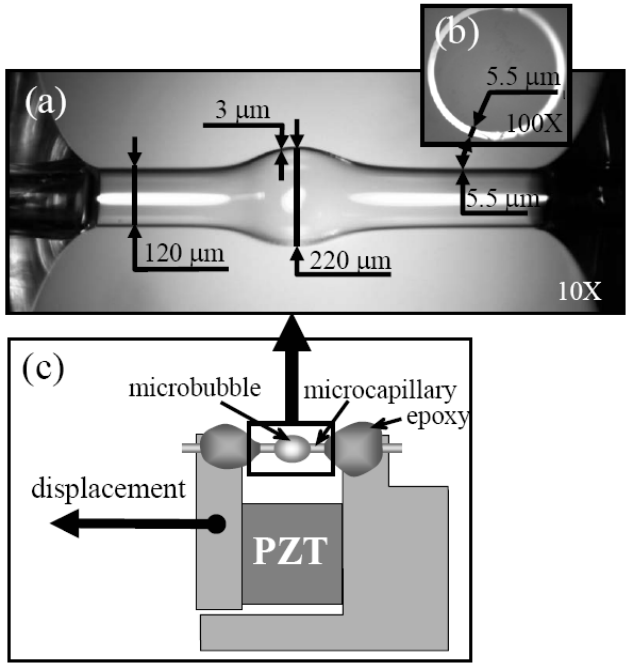

Fig. 2. (a) – Optical microscope picture of the MBR used in the experiment. (b) – Optical microscope picture of the cross-section of the capillary from which the MBR was created. (c) – Illustration of the setup used for the MBR stretching.

The MBR capillary ends are glued to the PZT stages as illustrated in Fig. 2(a) and (c). Light is coupled into the MBR with a biconical taper that is fabricated from a standard single mode fiber by the indirect $CO_2$ laser heating method [26]. The microfiber waist of this taper has a diameter of ~ 1.7 μm and length ~ 3.5 mm. The microfiber is positioned in direct contact with the microbubble along its equator. The taper is slightly relaxed in order to avoid the displacement of the microfiber with respect to the MBR in the process of stretching. The ends of the taper are connected to the JDS Uniphase tunable laser source and detector. The measured Q-factor of the observed resonances, limited by the detector resolution, 3 pm, exceeds $10^6$. Fig. 3(a) shows the spectral fragments taken in the wavelength interval 1554±4 nm (equal to ~ 3.5 $\Delta\lambda_{FSR}^{(azim)}$) with 2.5 μm steps of the PZT displacement. More detailed picture of the spectral evolution is given by Fig. 3(b), which shows the magnified spectral profiles (now, after each 0.5 μm step of the PZT displacement) of the region outlined by a bold rectangular in Fig. 3(a). In our experiment, the translation range of the PZT was limited by 30 μm. This restriction did not allow us to determine the actual tunable range of the MBR (with the same PZT, the tunability can be increased by decreasing the distance between the epoxy and the microbubble, see Fig. 2(a), (c)). Nevertheless, from Fig. 3(a), the available 30 μm stretching corresponds to the resonance wavelength shift of ~ 5.5 nm, which results in the tunability FOM $\Delta\lambda/\Delta\lambda_{FSR} \approx 2.2$, the record value achieved for the high Q-factor WGM microresonators.

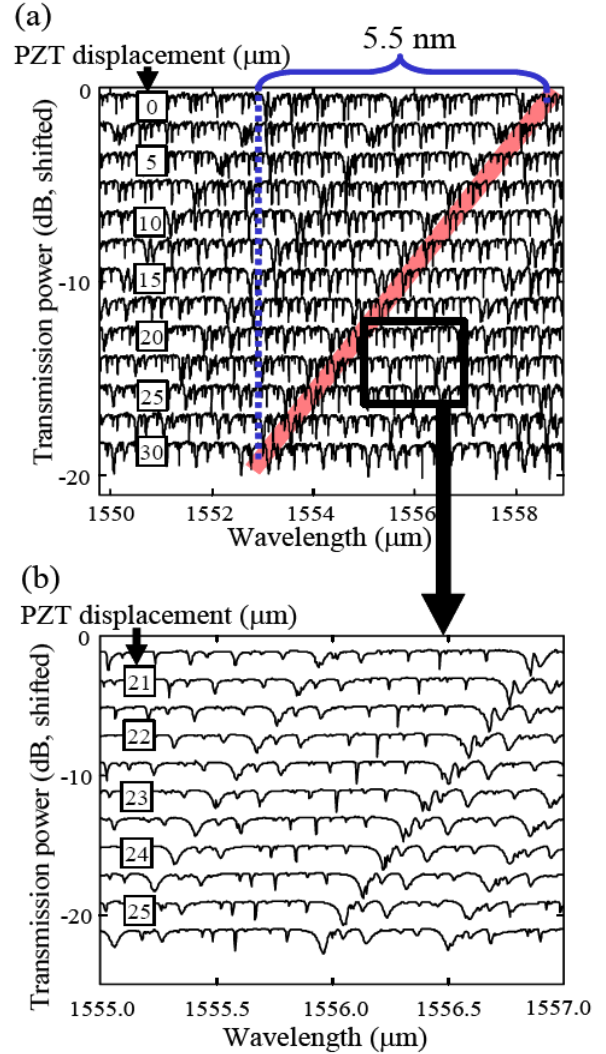

Fig. 3. (a) – Transmission spectra of the MBR for different PZT displacements. (a) – PZT displacement step is 2.5 μm. (b) – PZT displacement step is 0.5 μm.

In summary, we have demonstrated a fully tunable silica MBR having a tuning range more than two times larger than the azimuthal FSR of this resonator. This is the largest tuning range ever demonstrated for the high Q-factor microcavities (microspheres, microspheroids, microtoroids, etc.) significantly exceeding the record result obtained previously. Tunability of an MBR is accomplished with a much smaller stretching force than that needed for a similar bulk microresonator. The translation range limit of our PZT did not allow us to determine the actual tunable range of the MBR, which, according to our theoretical estimates, can be considerably greater than that we have demonstrated. A similar approach allows to fabricate much smaller MBRs from silica capillary fibers with smaller diameters.

The authors are grateful to D. J. DiGiovanni and V. S. Ilchenko for useful discussions.

## References

1. A. B. Matsko and V. S. Ilchenko, "Optical resonators with whispering gallery modes I: basics," IEEE J. Sel. Top. Quantum Electron. **12**, 3-14 (2006).
2. V. S. Ilchenko, P. S. Volikov, V. L. Velichansky, F. Treussart, V. Lefevre-Seguin, J.-M. Raimond, S. Haroche, "Strain-tunable high-Q optical microsphere resonator", Opt. Comm. **145**, 86-90 (1998).
3. W. von Klitzing, R. Long, V. S. Ilchenko, J. Hare, and V. Lefèvre-Seguin, "Frequency tuning of the whispering-gallery modes of silica microspheres for cavity quantum electrodynamics and spectroscopy," Opt. Lett. **26**, 166-168 (2001).
4. M. Cai and K. Vahala, "Highly efficient hybrid fiber taper coupled microsphere laser," Opt. Lett. **26**, 884-886 (2001).
5. H. C. Tapalian, J.-P. Laine, and P. A. Lane, "Thermooptical Switches Using Coated Microsphere Resonators," IEEE Photon. Technol. Lett. **14**, 1118-1120 (2002).
6. D. K. Armani, T. J. Kippenberg, S. M. Spillane, and K. J. Vahala, "Ultra-high-Q toroid microcavity on a chip," Nature **421**, 925-928 (2003).
7. D. Armani, B. Min, A. Martin, and K. J. Vahala, "Electrical thermo-optic tuning of ultrahigh-Q microtoroid resonators," Appl. Phys. Lett. **85**, 5439-5441 (2004).
8. A.A. Savchenkov, V.S. Ilchenko, A.B. Matsko and L. Maleki, "Tunable filter based on whispering gallery modes," Electron. Lett. **39**, 389-390 (2003).
9. V.S. Ilchenko, A.A. Savchenkov, A.B. Matsko and L. Maleki, "Tunability and synthetic lineshapes in high-Q optical whispering gallery modes," Proceedings of SPIE **4969**, 195-216 (2003).
10. M. Sumetsky, "Whispering-gallery-bottle microcavities: the three-dimensional etalon," Opt. Lett. **29**, 8-10 (2004).
11. G. S. Murugan, J. S. Wilkinson, and M. N. Zervas, "Selective excitation of whispering gallery modes in a novel bottle microresonator," Opt. Express **17**, 11916-11925 (2009).
12. M. Pöllinger, D. O'Shea, F. Warken, and A. Rauschenbeutel, "Ultra-high-Q tunable whispering-gallery-mode microresonator," Phys. Rev. Lett. **103**, Art. 053901 (2009).
13. M. Sumetsky, "Whispering gallery modes in a microfiber coil with an *n*-fold helical symmetry: classical dynamics, stochasticity, long period gratings, and wave parametric resonance," Opt. Express **18**, 2413-2425 (2010).
14. Here we exclude the situation when the resonances are degenerated.
15. W. C. L. Hopman, K. O. van der Werf, A. J. F. Hollink, W. Bogaerts, V. Subramaniam, and R. M. de Ridder, "Nano-mechanical tuning and imaging of a photonic crystal micro-cavity resonance," Opt. Express **14**, 8745-8752 (2006).
16. I. W. Frank, P. B. Deotare, M. W. McCutcheon, M. Loncar, "Dynamically reconfigurable photonic crystal nanobeam cavities," http://arxiv.org/abs/0909.2278.
17. A. Guarino, G. Poberaj, D. Rezzonico, R. Degl'Innocenti, and P. Gunter, "Electro-optically tunable microring resonators in lithium niobate," Nature Photonics **1**, 407–410 (2007).
18. R. Shamai and U. Levy, "On chip tunable micro ring resonator actuated by electrowetting," Opt. Express **17**, 1116-1125 (2009).
19. H. Mabuchi and H. J. Kimble, "Atom galleries for whispering atoms: binding atoms in stable orbits around an optical resonator," Opt. Lett. **19**, 749-751 (1994).
20. D. J. Norris, M. Kuwata-Gonokami, and W. E. Moerner, "Excitation of a single molecule on the surface of a spherical microcavity," Appl. Phys. Lett. **71**, 297-299 (1997).
21. J. R. Buck and H. J. Kimble, "Optimal sizes of dielectric microspheres for cavity QED with strong coupling," Phys. Rev. A **67**, Art. 033806 (2003).
22. M. Sumetsky, Y. Dulashko, and R. S. Windeler, "Optical microbubble resonator," Opt. Lett. **35**, 898-900 (2010).
23. Y. Hiramatsu and Y. Oka, "Determination of the tensile strength of rock by a compression test of an irregular test piece," Int. J. Rock Mech. Min. Sci. **3**, 89-99 (1966).
24. S. Timoshenko and S. Woinowsky-Krieger, "Theory of plates and shells," (McGraw-Hill Book Company, Inc, New York, 1959).
25. C. R. Kurkjian, J. T. Krause, and M. J. Matthewson, "Strength and fatigue of silica optical fibers," J. Lightwave Tech. **7**, 1360-1370 (1989).
26. M. Sumetsky, Y. Dulashko, and A. Hale, "Fabrication and study of bent and coiled free silica nanowires: Self-coupling microloop optical interferometer," Opt. Express **12**, 3521-3531 (2004).